\documentclass[11pt,a4paper,nofootinbib]{article} 
 \pdfoutput=1
\usepackage[utf8]{inputenc}
\usepackage{jheppub}
\usepackage{floatflt,rotate, cancel}
\usepackage{mathrsfs}
\usepackage{amssymb, amsmath}
\usepackage{color}
\usepackage{graphicx}
\usepackage{subcaption, caption}
\usepackage{multirow}
\usepackage{float}
\usepackage{pstricks}
\usepackage{cleveref}
\usepackage{soul}

\title{\boldmath A closer look at dark matter production in exponential growth scenarios}

\author{Disha Bhatia}
\affiliation{Instituto de Fisica, Universidade de São Paulo, C.P. 66.318, 05315-970 São Paulo, Brazil}

\emailAdd{dishabhatia@usp.br}

\abstract{
We investigate a recently proposed non-thermal mechanism for dark matter production, in which a small initial dark matter ($\chi$) number density undergoes exponential growth through scatterings with bath particles ($\phi$) in the early universe ($\chi \phi \to \chi \chi$). The process ends when the scattering rate becomes Boltzmann suppressed. 
The analysis, in literature, is
performed on the simplifying assumption of the dark matter phase space tracing the equilibrium
distribution of either standard model or a hidden sector bath. Owing to the non-thermal nature
of the production mechanism, this assumption may not necessarily hold.
In this work, we test the validity of this assumption by numerically solving the unintegrated Boltzmann equation for the dark matter distribution. Our results, independent of the initial conditions, show that after exponential growth ceases, the dark matter distribution exhibits equilibrium-like behaviour at low comoving momentum, especially for higher couplings. While full kinetic equilibrium-like behaviour is not reached across all momentum modes, the scaled equilibrium approximation provides reasonable estimates for the dark matter abundance. For more accurate results, however, the full unintegrated Boltzmann equation must be solved.
}

\keywords{Cosmology of Theories beyond the SM, Beyond Standard Model}

\DeclareUnicodeCharacter{2212}{-}

\begin{document}

\maketitle

\section{Introduction}
The details related to the
production mechanism of dark matter (DM) in the early universe and 
its detection at the experiments constitute 
two of the most pressing questions in the field of dark matter.
The freeze-out mechanism, where the dark matter is 
produced thermally 
(just like standard model particles), is thought to be 
one of the promising mechanisms to explain dark matter 
abundance~\cite{Kolb:1990vq}. This is because, in this scenario, the couplings 
required to explain relic density can also be potentially probed at the experiments~\cite{Feng:2010gw}.
However, due to the null results of the detection of dark matter so far at various experiments~\cite{XENON:2022ltv, Abulaiti:2799299,PerezAdan:2023rsl} 
it has become necessary
to study different kinds of production mechanisms besides standard thermal freeze-out~\cite{Baer:2014eja}.
Gravitational production of dark matter~\cite{Parker:1969au,Parker:1971pt,Ford:2021syk}, freeze-in~\cite{Hall:2009bx},
strongly interacting massive particles~\cite{Hochberg:2014dra}, hidden sector freeze-out~\cite{Pospelov:2008jk,Feng:2008mu} etc. 
are few of the mechanisms which can explain the non-observation of
particle nature of dark matter at the terrestrial experiments.

A new non-thermal production mechanism has recently been proposed which leads to the exponential growth 
of the dark matter ($\chi$) with the expansion of the universe~\cite{Bringmann:2021tjr,Hryczuk:2021qtz}.
The mechanism can be realised in scenarios where dark matter interacts with the bath particle ($\phi$) 
through the scatterings of the kind: $\chi + \phi \to \chi + \chi$.
These interactions can naturally arise in models with mass-mixings 
or scalar self-interactions~\cite{DEramo:2010keq,Bringmann:2021tjr,Hryczuk:2021qtz,Bringmann:2022aim}. 
The exponential growth at some stage of the expansion must stop otherwise it leads to 
the over-production of dark matter. 
For the case when mass of $\phi$ is larger than the dark matter 
mass, the production of dark matter stops once $\phi$ becomes non-relativistic.
In the opposite case where dark matter is heavier than the bath particle, the production stops 
when $\phi$ has insufficient energy to generate dark matter particles. 
The phenomena of exponential growth is also known as semi-production in the literature~\cite{Hryczuk:2021qtz}
because it is complimentary to the dark matter production via semi-annihilation mechanism~\cite{DEramo:2010keq}.
This new mechanism of dark matter production proceeds through the scatterings between 
dark matter and bath particles, therefore it cannot solely generate full dark matter 
abundance. For this mechanism to be active, there must necessarily be some non-zero dark matter density present 
in the early universe before the onset of exponential production. 
Such initial density can be generated by multiple processes for example, 
freeze-in, gravitational 
production of dark matter, inflaton decay etc.

For any production mechanism, thermal or non-thermal, relic density of dark matter today 
can be computed by solving the Boltzmann equation (B.E.) in the expanding universe
which determines the evolution 
of the dark matter phase space distribution ($f$)~\cite{Gondolo:1990dk}. 
As Boltzmann equation is an integrodifferential equation and 
solving it for generic cases is highly non-trivial~\cite{Binder:2017rgn}.
There are broadly two ways to solve for Boltzmann equation.
The first way is to solve full unintegrated Boltzmann equation i.e. $L[f] = C[f]$
where $L$ is the Liouville's operator, and $C$ is the collision operator
which encapsulates both elastic and inelastic scatterings of 
dark matter. To obtain solutions using this method is often complicated in absence of 
any simplifications owing to the highly non-linear nature of the Boltzmann equation. The second method is to solve for
coupled differential equation for different moments\footnote{Moments of BE can be determined by integrating the 
weighted BE equation over full phase space and performing spin summation. For example, nth moment of BE is 
defined as $\int \frac{d^3p}{(2\pi)^3} g \; p^n f(p,t)$, where $p^n$ is the weight 
factor and $g$ is the spin degrees of freedom. The zeroth and the second moment equation physically represent the evolution  
of the number density and energy/temperature with time respectively.}. 
In the scenarios where dark matter is thermally produced, 
the solution to relic density is relatively simple. 
This is because in such cases dark matter is 
at least in kinetic equilibrium with the either SM-bath or some hidden sector bath,
and its phase space distribution 
can be approximated to follow the equilibrium pattern~\cite{Gondolo:1990dk,Binder:2017rgn} 
i.e. 
\begin{equation}
f_\chi(p,t) = A(t) f_\chi^{\rm eq}(p,t) \;, \quad \text{where} \quad A(t) = n_\chi(t)/n_\chi^{\rm eq}(t) \;.
\label{eqn:distribution-fun}
\end{equation}
In particular, it becomes sufficient, in above case, to solve the zeroth-moment or/and second-moment Boltzmann equation depending on whether we have equilibrium with 
the standard model bath or the hidden sector. 

The dark matter abundance through the exponential growth mechanism in the literature has been 
determined by using the simplified assumption where the dark matter distribution function 
traces an equilibrium 
distribution as in eqn.~(\ref{eqn:distribution-fun}) with a temperature 
either of the standard model bath~\cite{Bringmann:2021tjr} or some hidden sector bath~\cite{Hryczuk:2021qtz, Bringmann:2022aim}.
The consideration 
regarding the distribution function following 
the equilibrium behaviour, although is well 
justified for thermal or nearly thermal scenarios~\cite{Binder:2017rgn}, may not always hold for 
purely non-thermal production mechanisms. 
In the present work, we depart from the assumption 
and directly solve the unintegrated Boltzmann equation numerically and determine the phase space distribution function of dark matter. 
Although solving for a full BE equation in itself is a complicated problem~\cite{Binder:2017rgn},
it simplifies a bit in our case due to the negligible amount of dark matter number density
in comparison with the number density of the bath particle. Due to this, the backward scattering 
process ($\chi \chi \to \chi \phi$) can be safely neglected and
Boltzmann equation reduces to a coupled linear partial differential equation. 
The computation can be further simplified in the comoving frame\footnote{There is an additional advantage of solving the Boltzmann equation in the comoving frame, as the distribution function obtained after dark matter abundance saturation does not redshift with the expansion of the universe and remains constant until the present epoch. This, however, holds specifically for adiabatic expansions. In our work, we assume that entropy remains conserved, resulting in an adiabatic expansion.} of the dark matter particle~\cite{DAgnolo:2017dbv} where the system of coupled partial differential equations reduces to a system of coupled ordinary differential equations.

In our analysis, we assume that the initial population of dark matter is either generated by the out-of-equilibrium decay of a heavy particle or produced by a process that leads to an equilibrated distribution of dark matter in the early universe. Using these initial conditions, we solve the system of coupled first-order ordinary differential equations to determine the effect of exponential growth on the dark matter distribution and its abundance.
Our numerical results show that, independent of the initial conditions, after exponential growth has ceased, the final dark matter distribution exhibits equilibrium-like behaviour at low comoving momentum, with this behaviour becoming more pronounced for higher couplings\footnote{By equilibrium-like behaviour, we imply that the dark matter distribution follows a scaled equilibrium function as given in eqn.~(\ref{eqn:distribution-fun}). It is still far from attaining chemical equilibrium with the thermal bath.}. This occurs because, for low momentum and large couplings, the mean free path of the dark matter particles is short. As a result, dark matter remains in the thermal bath longer, leading to more scatterings. In contrast, for larger momenta, the mean free path is longer, reducing scatterings with thermal particles and resulting in later thermalization.
The dark matter distribution does not reach full kinetic equilibrium-like behaviour\footnote{We mention kinetic and not chemical, as the couplings required to establish chemical equilibrium are still too low.} across the entire momentum range, as the coupling required to establish the correct abundance is lower than what is needed to achieve kinetic equilibrium. However, equilibrium behaviour is attained for low-momentum modes. Therefore, while the approximation of dark matter tracing the equilibrium distribution during the exponential production phase is not entirely accurate, it provides a reasonable estimate for approximate results. 
For practical purposes, determining the dark matter abundance using the scaled equilibrium assumption~\cite{Bringmann:2021tjr, Hryczuk:2021qtz} is effective, as solving the unintegrated Boltzmann equation is computationally complex. Moreover, the couplings needed to satisfy the dark matter abundance in both methods fall within a similar range.

The paper is organized as follows: In section~\ref{sec:general}, we discuss the unintegrated Boltzmann equation to analyse scenarios of dark matter 
production via the exponential growth.  
In section~\ref{sec:specific-examples}, we examine two initial condition scenarios for obtaining the final dark matter distribution function. 
In the sub-section~\ref{sec:scaledeq},
we consider a model independent approach and consider the initial dark matter distribution function to be 
the scaled equilibrium distribution without specifying the mechanism generating dark matter's initial abundance, and in sub-section~\ref{sec:inflation}, we consider that dark matter initial abundance 
to be generated due to the out-of-equilibrium decay of the heavy particle. Finally, in section~\ref{sec:summary}, 
we summarize our findings and present the concluding remarks. In appendix~\ref{sec:appendix}, we compute the dark matter abundance using the equilibrium 
approximation, and discuss phase diagrams to show complementarity between exponential production mechanism and semi-annihilations.

\section{The unintegrated Boltzmann equation for exponential production}
\label{sec:general}

\begin{figure}
    \centering
    \includegraphics[width=0.65\linewidth]{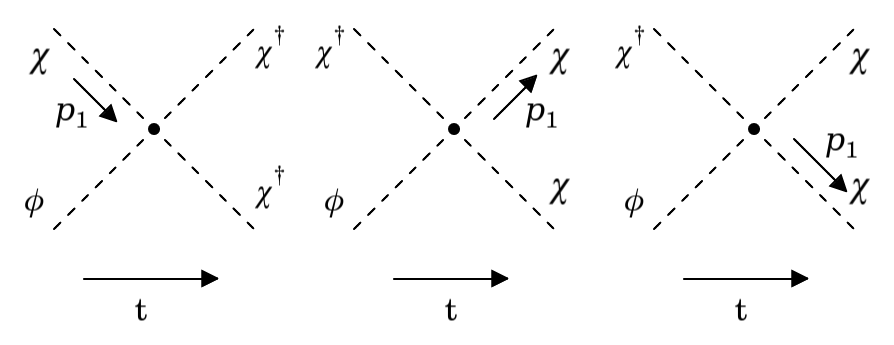}
    \caption{Scatterings between dark matter ($\chi$) and bath particle ($\phi$) leading to the 
    exponential growth. Diagram (a) leads to annihilation of the dark matter with momentum $p_1$, and 
    diagrams (b) and (c) lead to creation of dark matter particle with momentum $p_1$.}
    \label{fig:feynman-dia}
\end{figure}
\noindent In this section, we discuss the broad 
features of dark matter ($\chi$) production through 
exponential growth mechanism in the early universe. This 
production mechanism is active if there are scatterings of 
dark matter with the bath particle ($\phi$) of the kind: $\chi + \phi \to \chi + \chi$
to produce more dark matter. For this scattering process to begin, we necessarily require
presence of some initial non-zero dark matter abundance making the process is 
initial condition dependent. As dark matter is not in equilibrium with any thermal bath, hence 
to determine its abundance one needs to solve the unintegrated Boltzmann equation. To 
solve this equation, we require two inputs: 
\begin{enumerate}
\item First input is regarding details of the model 
which initiates the dark matter exponential growth. We consider the toy model of 
a scalar dark matter charged under a  $Z_3$ symmetry and coupled with the bath particle\footnote{
The particle $\phi$ is standard model singlet. It is thermally coupled with the standard model plasma through its interactions 
with the Higgs $\frac{\lambda_{\rm S}}{2!} \mu H^\dagger H \phi   + \frac{\lambda_{\rm S^\prime}}{2!} H^\dagger H \phi^2 $
The scalar coupling $\lambda_{S}$ and $\lambda_{S^\prime}$ 
can be suitably chosen such that $\phi$ remains in thermal equilibrium 
with the standard model bath in the early universe,
and does not receive stringent constraints from the Higgs data~\cite{ATLAS:2019nkf,ATLAS:2022vkf}} $\phi$ as 
$\mathcal{L}_{\rm int} = \frac{\lambda_{\rm{exp}}}{3!} \chi^3 \phi $ + h.c. ~\cite{Bringmann:2021tjr,Hryczuk:2021qtz}.
This interaction will give rise to scatterings
\begin{equation}
\chi + \phi \to \chi^\dagger + \chi^\dagger \quad \text{and} \quad \chi^\dagger + \phi \to \chi + \chi\;,
\end{equation}
required for 
exponential production.  Note that the mass of $\phi$ should be less than three times the mass of dark matter to 
avoid direct production of dark matter through the decays of $\phi$.
\item Since Boltzmann equation is a differential equation, to solve it we require the knowledge of initial 
conditions, specifically the initial dark matter distribution function. This input is regarding 
the information of the mechanism which generates
the initial abundance of the dark matter. There are large number of possibilities here, 
the initial abundance could have been generated by gravitational production of dark matter, 
freeze-in process etc.. In this section, we remain as generic as possible in our discussion
and follow a model-independent approach regarding the initial abundance of dark matter. In section~\ref{sec:specific-examples}, 
we consider few initial conditions to obtain the final dark matter distribution.

\end{enumerate}

To solve the unintegrated Boltzmann equation i.e. $\mathcal{L}[f] = \mathcal{C}[f]$ for the distribution function, 
we need to consider the scatterings which annihilates and creates dark matter $\chi$ of a particular momentum, 
and then write equations for all such possible allowed momenta. 
Let us consider the creation and annihilation of the dark matter with momentum $p_1$ at time t. This is displayed 
in figure~\ref{fig:feynman-dia}. The first diagram leads to annihilation of dark matter with momentum $p_1$, 
and the second and third diagram leads to creation of dark matter with momentum $p_1$ at time $t$. The Liouville and collision operator 
for this is given as: 
\begin{eqnarray}
  \mathcal{L}[f_\chi(p_1,t)] &=& \left(\frac{\partial}{\partial t} - H p\frac{\partial}{\partial p}\right) f_\chi(p_1,t)\;,\nonumber \\  
 \mathcal{C}[f_\chi(p_1,t)] &=&  \textcolor{red}{-}\frac{1}{2 E_1\; S}   \int \frac{d^3 p_2}{(2\pi)^3 2 E_{2}} \frac{d^3 p_3}{(2\pi)^3 2 E_3}  \frac{d^3 p_4}{(2\pi)^3 2 E_{4}}  
  \times  (2\pi)^4 \delta^4(p_1 + p_2 -p_3 -p_4)  \; \nonumber \\ 
  && \times  |M|^2_{\chi \phi \to \chi^\dagger \chi^\dagger}  \times \bigg[ f_\chi(p_1, t) f^{\rm eq}_\phi(p_2, t) \left( 1 \pm f_{\chi^\dagger}(p_3, t) \right) \left( 1 \pm f_{\chi^\dagger}(p_4, t) \right) \nonumber \\
  && \quad - f_{\chi^\dagger}(p_3, t) f_{\chi^\dagger}(p_4, t) \left( 1 \pm f_\chi(p_1, t) \right) \left( 1 \pm f_\phi^{{\rm eq}}(p_2, t)\right) \bigg]\; \nonumber \\
  && \textcolor{red}{+2} \; \frac{1}{2 E_1\; S}   \int \frac{d^3 p_2^\prime}{(2\pi)^3 2 E_{2}^\prime} \frac{d^3 p_3^\prime}{(2\pi)^3 2 E_3^\prime}  \frac{d^3 p_4^\prime}{(2\pi)^3 2 E_{4}^\prime}  
  \times  (2\pi)^4 \delta^4(p_3^\prime + p_2^\prime -p_1 -p_4^\prime)   \; \nonumber \\ 
  && \times |M|^2_{\chi^\dagger \phi \to \chi \chi} \times \bigg[ f_{\chi^\dagger}(p_3^\prime, t) f^{\rm eq}_\phi(p_2^\prime, t) \left( 1 \pm f_\chi(p_1, t) \right) \left( 1 \pm f_\chi(p_4^\prime, t) \right) \nonumber \\
  && \quad - f_\chi(p_1, t) f_\chi(p_4^\prime, t) \left( 1 \pm f_{\chi^\dagger}(p_3^\prime, t) \right) \left( 1 \pm f_\phi^{{\rm eq}}(p_2^\prime, t)\right) \bigg]\;.
    \label{eqn:collisionRG}
\end{eqnarray}
The above equation is an infinitely coupled integrodifferential equation. 
Note that collision operator in general constitutes all possible allowed scatterings which can create or annihilate dark matter. 
We assume that the dominant contribution arise due to the exponential production, and correspondingly neglect all other scatterings in our analysis. 
A similar equation like eqn.~(\ref{eqn:collisionRG}) can be written for the evolution of the distribution function for $\chi^\dagger$ i.e. $f_{\chi^\dagger}(p_1,t)$. 
We solve the Boltzmann equation for 
the scenarios where there is no asymmetry between $\chi$ and $\chi^\dagger$ and the chemical potential is zero. 
In this case, the distribution function of particle and anti-particle
dark matter is identical i.e. $f_\chi(p,t) \equiv f_{\chi^\dagger}(p,t)$. 
The matrix amplitude squared for $\chi + \phi \to \chi^\dagger + \chi^\dagger$ and $\chi^\dagger + \phi \to \chi + \chi$ scatterings are also 
identical. Note that $S$ in eqn.~(\ref{eqn:collisionRG}) refers to the symmetry factor\footnote{The symmetry factor is two for the toy model considered in our paper.} and the factors of \textcolor{red}{$-1$} and \textcolor{red}{$+2$} are highlighted in red in eqn.~(\ref{eqn:collisionRG}) is to show creation and annihilation of the dark matter 
of momentum $p_1$ at time $t$.

We can make some simplifications in the collision term as the dark matter is produced in an out-of-equilibrium 
process and its number density is much smaller than that of the bath particle ($\phi$). 
In exponential production, even though the dark matter number density increases very sharply, 
it never reaches equilibrium density. Consequently, the rate of interaction governing dark matter production is always smaller than the 
rate of expansion of universe. Hence, we can safely neglect the backward scattering process
along with the Pauli Blocking and Bose enhancement factors in our computations. 
With these simplifications, the collision term tracing the evolution of dark matter 
moving with momentum $p_1$ at time $t$ reduces to a linear equation in dark matter distribution function and is 
infinitely coupled to other momenta i.e. $\int d p_3^\prime \; f_\chi(p_3^\prime,t)$.  
The Boltzmann equation becomes a system of coupled linear partial differential equations to solve.
For our analysis, we write the collision term in two parts, the first part $\mathcal{C}_{\rm I}[\textcolor{blue}{f_\chi(p_1,t)}]$ is 
responsible for depletion of dark matter of momentum $p_1$ i.e. $\chi(p_1) \phi(p_2) \to \chi(p_3) \chi(p_4)$, and the second part $\mathcal{C}_{\rm II}[\textcolor{blue}{f_\chi(p_1,t)}]$ 
is responsible for the production of dark matter of momentum $p_1$ at time $t$ i.e $\chi(p_3^\prime) \phi(p_2^\prime) \to \chi(p_1) \chi(p_4^\prime)$. The simplified collision term is given as: 
\begin{eqnarray}
 \mathcal{C}[\textcolor{blue}{f_\chi(p_1,t)}] &=& \mathcal{C}_{\rm I}[\textcolor{blue}{f_\chi(p_1,t)}] + \mathcal{C}_{\rm II}[\textcolor{blue}{f_\chi(p_1,t)}]  \quad \quad \text{where }\nonumber \\ 
  \mathcal{C}_{\rm I}[\textcolor{blue}{f_\chi(p_1,t)}]&=& \textcolor{red}{-}\frac{1}{2 E_1\; S}   \int \frac{d^3 p_2}{(2\pi)^3 2 E_{2}} \times \textcolor{blue}{f_\chi(p_1, t)} f^{\rm eq}_\phi(p_2, t)  
   \; \nonumber \\ 
  && \times \int \frac{d^3 p_3}{(2\pi)^3 2 E_3}  \frac{d^3 p_4}{(2\pi)^3 2 E_{4}}   (2\pi)^4 \delta^4(p_1 + p_2 -p_3 -p_4)  \times  |M|^2_{\chi \phi \to \chi^\dagger \chi^\dagger}   \nonumber \\
 \mathcal{C}_{\rm II}[\textcolor{blue}{f_\chi(p_1,t)}] &=& \textcolor{red}{+2} \; \frac{1}{2 E_1\; S}   \int \frac{d^3 p_2^\prime}{(2\pi)^3 2 E_{2}^\prime} \frac{d^3 p_3^\prime}{(2\pi)^3 2 E_3^\prime}  \frac{d^3 p_4^\prime}{(2\pi)^3 2 E_{4}^\prime}  
   \; \nonumber \\ 
  && \times  (2\pi)^4 \delta^4(p_3^\prime + p_2^\prime -p_1 -p_4^\prime)   \times |M|^2_{\chi^\dagger \phi \to \chi \chi} \times \textcolor{blue}{f_{\chi}(p_3^\prime, t)} f^{\rm eq}_\phi(p_2^\prime, t)\;   \;.
\label{eqn:collision-operator}
\end{eqnarray}
Note that here we have highlighted dark matter distribution function in blue, to show how and where it 
exactly appears in the collision terms.
Note that for each collision term, there are total five 
integrals to be performed. For the collision term $\mathcal{C}_{\rm I}[\textcolor{blue}{f_\chi(p_1,t)}]$, we can perform the integrals over $p_3$ and $p_4$ in the center of 
mass frame, as the distribution functions do not depend explicitly on $p_3$ and $p_4$. Eventually we remain with two integrals to perform i.e. 
\begin{eqnarray}
 \mathcal{C}_{\rm I}[\textcolor{blue}{f_\chi(p_1,t)}] & = &   {-}\frac{|\lambda_{\rm exp}|^2}{16 (2 \pi)^3 E_1\; S} \textcolor{blue}{f_\chi(p_1,t)} \int dp_2 \;  d \cos{\theta} \frac{p_2^2 }{E_2} f_\phi(p_2,t) \sqrt{\frac{s-4 m_\chi^2}{s}} \\ 
 && \left( \text{where} \quad s = m_\chi^2 + m_\phi^2 + 2 E_1 E_2 - 2 p_1 p_2 \cos \theta  \right) \nonumber 
\end{eqnarray}
For the second collision term, $\mathcal{C}_{\rm II}[\textcolor{blue}{f_\chi(p_1,t)}]$ resulting in production of dark matter, 
the integrals are performed in the lab frame where we choose $p_1$ to be oriented along the z-axis, $p_4^\prime$ to be at an angle $\theta_1$ from the z-axis, and $p_3^\prime$ at 
angle $\theta_2$ with z-axis and lying in a different plane\footnote{$\overrightarrow{p_1} = (0,0,p_1)$, $\overrightarrow{p_4^\prime} = (p_4^\prime \sin \theta_1,0,p_4^\prime \cos \theta_1)$, 
$\overrightarrow{p_3^\prime} = (p_3^\prime \sin \theta_2 \cos\phi, p_3^\prime \sin \theta_2 \sin \phi, p_3^\prime \cos\theta_2)$ and $\overrightarrow{p_2^\prime} = \overrightarrow{p_1} + \overrightarrow{p_3^\prime} - \overrightarrow{p_4^\prime}$}.
The collision term becomes:
\begin{eqnarray}
    \mathcal{C}_{\rm II}[\textcolor{blue}{f_\chi(p_1,t)}] & = & \frac{|\lambda_{\rm exp}|^2}{8 (2 \pi)^4 E_1\; S} \int d \cos{\theta_1} \;  d \cos{\theta_2} \; d\phi \; d p_4^\prime \; \frac{{p_4^\prime}^2}{E_4^\prime} \textcolor{blue}{f_\chi(p_3^\prime,t)} f_\phi(p_2^\prime,t) \nonumber \\
    && \times \frac{{p_3^\prime}^2}{\left[p_3^\prime (E_1 + E_4^\prime) - E_3^\prime \left( \left(p_1 + p_4^\prime \cos\theta_1\right) \cos\theta_2 + p_4^\prime \cos\phi \sin \theta_1 \sin \theta_2 \right) \right]} \\
    && \left(\text{where} \; p_3^\prime \; \text{and} \; p_4^\prime \;  \text{are determined using energy-momentum conservation.}\right) \nonumber
\end{eqnarray}
In this collision term, there are four integrals to compute, as the fifth integral over azimuthal angle is trivial. This part 
of the collisions where dark matter of momentum $p_1$ is created is actually infinitely coupled to other Boltzmann equations of different momenta i.e. $\int d p_3^\prime \; \textcolor{blue}{f_\chi(p_3^\prime,t)}$.  
Hence we have a system of coupled linear partial differential equations to solve. The Boltzmann equation can be furthermore simplified by going to the comoving frame~\cite{DAgnolo:2017dbv} of the 
initial dark matter candidate which scatters off the bath particle.  
As the comoving momentum $q$ does not red-shift with the expansion of the universe, the Liouville's 
operator acting on the dark matter distribution function simplifies i.e. $\mathcal{L}[f_\chi(q,t)] = \frac{\partial}{\partial t}[f_\chi(q,t)]$. 
The Boltzmann equation transforms to set of infinite ordinary differential equations for each comoving momenta ($q$).

Tracing the evolution of 
the dark matter distribution function becomes physical if we replace time variable by a dimensionless-variable
characterizing the temperature (T) of the standard model thermal bath. We express the distribution function of dark matter 
in terms of $x$ and comoving momentum $q$ which are defined as following:
\begin{equation}
     q \equiv \frac{p}{{[s(x)}]^{1/3}}  \;, \ x \equiv \frac{m_\phi}{T} \quad \;.
     \label{eqn:comoving}
\end{equation}
The comoving momenta is defined using conservation of entropy and does not red shift with the expansion of the universe. 
The variable $s(x)$ is the 
entropy density of the thermal bath at epoch $x$.  
Note that using the variables defined in eqn.~(\ref{eqn:comoving}), the zeroth moment of the phase space distribution viz. $f(q, x)$ 
physically represent the comoving number density of dark matter 
particle i.e. $Y_\chi(x) = n_\chi(x)/s(x)$. Using the above simplifications and transformations, 
the Liouville's and the collision operator 
in eqn.~(\ref{eqn:collisionRG}) can be expressed as:
\begin{eqnarray}
    \mathcal{L}[f_\chi(q_1,x)] &=& \frac{x H(x)}{h_{\rm eff}(x)}\frac{\partial f_\chi(q_1,x)}{\partial x} \;, \nonumber \\
   \mathcal{C}_{\rm I}[\textcolor{blue}{f_\chi(q_1,x)}] & = &   {-}\frac{|\lambda_{\rm exp}|^2}{16 (2 \pi)^3 E_1\; S} \textcolor{blue}{f_\chi(q_1,x)} s(x) \int dq_2 \;  d \cos{\theta} \frac{q_2^2 }{E_{q_2}} f_\phi(q_2,x) \sqrt{\frac{s-4 m_\chi^2}{s}} \nonumber \\
   \mathcal{C}_{\rm II}[\textcolor{blue}{f_\chi(q_1,t)}] & = & \frac{|\lambda_{\rm exp}|^2}{8 (2 \pi)^4 E_1\; S} s(x)^{\frac{4}{3}}\int d \cos{\theta_1} \;  d \cos{\theta_2} \; d\phi \; d q_4^\prime \; \frac{{q_4^\prime}^2}{E_{q_4}^\prime} \textcolor{blue}{f_\chi(q_3^\prime,x)} f_\phi(q_2^\prime,x) \nonumber \\
    &\times &  \frac{{q_3^\prime}^2}{\left[q_3^\prime (E_{q_1} + E_{q_4}^\prime) - E_{q_3}^\prime \left( \left(q_1 + q_4^\prime \cos\theta_1\right) \cos\theta_2 + q_4^\prime \cos\phi \sin \theta_1 \sin \theta_2 \right) \right]} \;.
 \label{eqn:BEOpLevel}
\end{eqnarray}
Here $h_{\rm eff}(x) = \left( 1 - \frac{1}{3} \frac{d \log{g_{*,s}(x)}}{d \log x}\right)$, $H(x)$
is the Hubble rate and $E_{q_i}$ is energy written in terms of comoving momentum.

We discretize the comoving momenta i.e. $f({q_i,x}) \equiv f_i(x)$
to solve the coupled differential equation effectively. Since the Boltzmann equation is linear in the distribution function, it can be expressed compactly
using the matrices: 
\begin{equation}
\frac{d}{dx} \begin{pmatrix}
f_1(x) \\
f_2(x) \\
\vdots \\
f_n(x)
\end{pmatrix} = A[x] \begin{pmatrix}
f_1(x) \\
f_2(x) \\
\vdots \\
f_n(x)
\end{pmatrix}
\label{eqn:coupledBE-numerical}
\end{equation}
Here $A[x]$ is a $n\times n$ matrix determined after solving the collision terms $\mathcal{C}_{\rm I}[\textcolor{black}{f_\chi(q_1,x)}]$ and $\mathcal{C}_{\rm I}[\textcolor{black}{f_\chi(q_1,x)}]$ . 
We determine $A[x]$ numerically at different intervals of $x$ varied in the range [0.1,10] where most of the dark matter production 
takes place. The comoving momentum is discretized in the range [0.01-10] with the step-size of few hundred thousand in the computation.
The chosen range for the comoving momentum is appropriate for our problem, as both the initial dark matter distribution and the bath particle distribution lie within this range. The stability of the integration is checked by running the code for higher step-sizes. The analysis has been performed 
using {\tt Mathematica-14}~\cite{Mathematica}. 

It can be seen from equation~(\ref{eqn:BEOpLevel}) that the solutions to the coupled Boltzmann equation for the distribution function of dark matter grow roughly exponentially when both $\chi$ and $\phi$ are relativistic. In this regime, $E_{q_i}$ scales as $s(x)^{1/3}$, causing the collision terms to scale as $s(x)^{1/3}/(x H(x))$, which remains essentially constant with respect to $x$.
In the non-relativistic limit, $E_{q_i}$ is proportional to the mass and is independent of $x$. In this case, the growth rate is slower and eventually halts as the process becomes Boltzmann suppressed.
It is important to note that not all momentum modes are simultaneously relativistic or non-relativistic for a given $x$. Therefore, when solving the unintegrated coupled Boltzmann equation for $f_\chi(q,x)$, some momentum modes will grow exponentially, while others will grow more slowly, at a rate less than exponential. As a result, the comoving number density, $Y_\chi(x)$, computed numerically by solving unintegrated Boltzmann equation, does not exhibit uniform exponential growth. This is shown in left panel of figure~\ref{fig:eq-distribution}. This contrasts with the case where the equilibrium assumption is applied, allowing $Y_\chi$ to be computed by solving the Boltzmann equation in closed form using the zeroth moment, as discussed in section~\ref{sec:appendix}. In this case, $Y_\chi(x)$ grows uniformly before its production becomes Boltzmann suppressed.

\section{Dark matter distribution after exponential production}
\label{sec:specific-examples}
As we discussed, in order to solve for the dark matter number density in the exponential production scenario, 
a finite and non-zero dark matter density must have been present initially. Therefore, to obtain the final distribution function and, 
correspondingly, the number density of dark matter, we need to know the initial distribution function of dark matter. This contrasts with solving only for the zeroth-moment Boltzmann equation 
for the case with scaled equilibrium assumption, where knowing the dark matter number density alone is sufficient.
In this section, we consider two examples of the initial conditions -- first we 
consider the case where dark matter initial distribution can be parametrized by eqn.~(\ref{eqn:distribution-fun}) at $x_0$. 
In the second example, we consider that dark matter was initially produced through the out-of-equilibrium decay of a heavy particle.  
We then solve the Boltzmann equations to determine the distribution of dark matter after the exponential growth phase has occurred. 

\subsection{Case a: Scaled equilibrium distribution as the initial distribution}
\label{sec:scaledeq}
In this section, we remain agnostic regarding the initial production of dark matter and consider a simple, model-independent approach 
that the dark matter distribution at some initial epoch is accidentally given as in eqn.~(\ref{eqn:distribution-fun}). This choice is similar
to the one made in reference~\cite{Bringmann:2021tjr}. However, in that case, the DM distribution is assumed to follow 
eqn.~(\ref{eqn:distribution-fun}) for all epochs. In this section, we assume eqn.~(\ref{eqn:distribution-fun}) only for the initial epoch 
and aim to test how dark matter distribution function evolves with the onset of exponential production.
After solving the unintegrated Boltzmann equation, we are interested in determining whether the dark matter distribution continues to follow the behaviour in eqn.~(\ref{eqn:distribution-fun}) 
or whether it diverges from it. 

We consider the initial epoch to be $x_0=0.1$. After substituting eqn.~(\ref{eqn:distribution-fun}) in eqn.~(\ref{eqn:coupledBE-numerical}), we obtain the 
final dark matter distribution function due to exponential production. 
\begin{figure}
    \centering
    \includegraphics[width=0.46\linewidth]{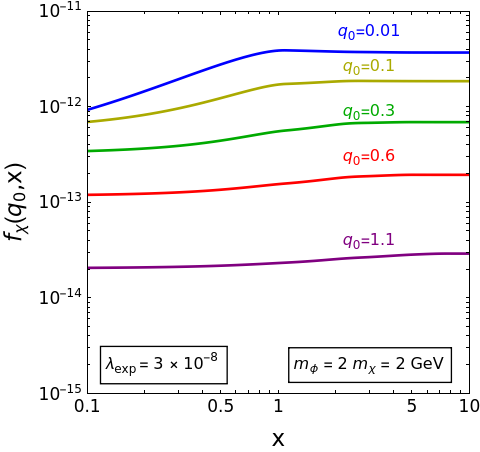}
     \includegraphics[width=0.45 \linewidth]{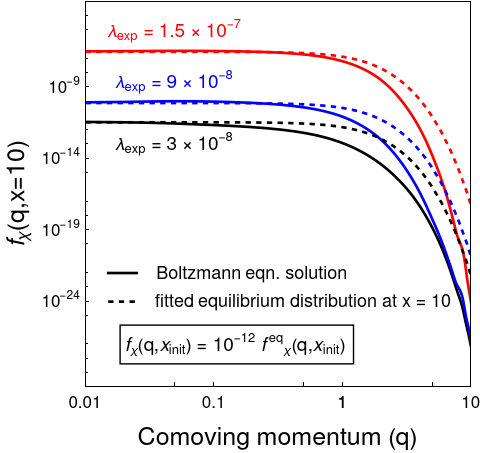} \\
    \caption{In the left panel, different modes of the dark matter distribution function are plotted against $x$ for a chosen coupling 
    of $\lambda_{\rm tr} = 3 \times 10^{-8}$. In the right panel, we plot the snapshot of the distribution function at $x=10$ where the 
    exponential growth has ceased. The solid lines are obtained by solving the unintegrated Boltzmann equation, while the dashed lines represent fits to the numerical results, with the fit function being the scaled equilibrium distribution as given in eqn.~(\ref{eqn:distribution-fun}). For the purpose of the analysis, the masses of dark matter and bath particle are chosen to be 1 and 2 GeV, respectively.}
    \label{fig:differentq}
\end{figure}
In the left panel of figure~\ref{fig:differentq}, we plot the dark matter distribution function as a function of 
$x$ for different values of comoving momenta. This plot has been made for the specific choices of dark matter and the bath particle 
mass as well the coupling parameter $\lambda_{\rm exp}$. However, the conclusions drawn are generic as long as we remain within the framework of exponential production and 
the considered initial conditions. It can be observed that the growth of different co-moving momenta exhibits distinct behaviours, with growth not being uniform across all modes. In particular, modes with smaller momenta grow much faster than those with higher momenta. This behaviour can be understood in terms of the mean free path: modes with larger momenta have longer mean free paths and consequently experience fewer scatterings. Additionally, high co-moving momentum modes freeze in at slightly later epoch than lower momentum modes, as they remain relativistic for a longer duration at a given epoch $x$.

In the right panel of figure~\ref{fig:differentq}, we plot the distribution of DM as a function of comoving momenta at the epoch $x=10$, where its 
production has nearly ceased due to the exponential production.
Starting from a scaled equilibrium distribution at $x=x_{\rm init}$, we observe that the distribution function obtained after solving the unintegrated Boltzmann equation (shown as solid lines) evolves differently from the scaled equilibrium behaviour. To show this, we compare our numerical results with the fitted equilibrium distribution at $x=10$ plotted in dashed lines\footnote{The fit is performed by taking the numerical results obtained at 
$x=10$ for each coupling and fitting them to a scaled equilibrium distribution function, as given in equation eqn.~(\ref{eqn:distribution-fun})}. Although the initial conditions are assumed to be accidentally following the equilibrium behaviour, the phenomena generating dark matter via exponential growth is intrinsically non-thermal in nature, thereby generating departure from equilibrium behaviour. In particular, the low momentum modes 
grow at a faster rate than high momentum modes as can be seen from the left panel of figure~\ref{fig:differentq}, causing a departure from the initial equilibrated behaviour. 
The differences are more pronounced for higher comoving momentum values and lower couplings. For lower comoving momentum, the distribution function closely traces the scaled distribution function. As we increase the coupling value, higher comoving momentum modes begin to follow the equilibrium behaviour which again can be understood in terms of the mean free path which depends inversely on the cross-section. A higher cross-section results in a shorter mean free path, meaning that dark matter interacts more with the medium particles and can achieve equilibrium more quickly. This behaviour is evident in the cases of higher couplings and low momentum. Conversely, for lower couplings and higher momentum, the mean free path of dark matter becomes larger, leading to fewer interactions with the bath particles. As a result, these modes do not exhibit equilibrium behaviour. Note that the equilibrated behaviour just refers to the kinetic equilibrium.

\begin{figure}
    \centering
    \includegraphics[width=0.47\linewidth]{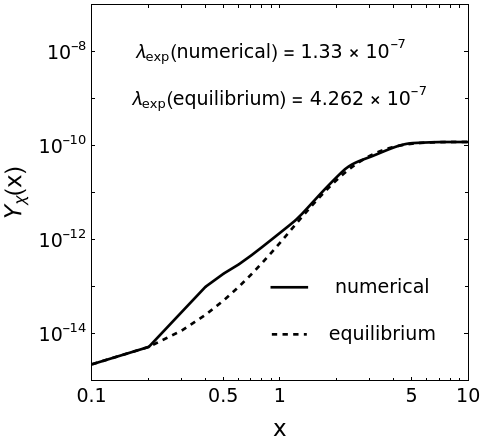}
    \includegraphics[width=0.45 \linewidth]{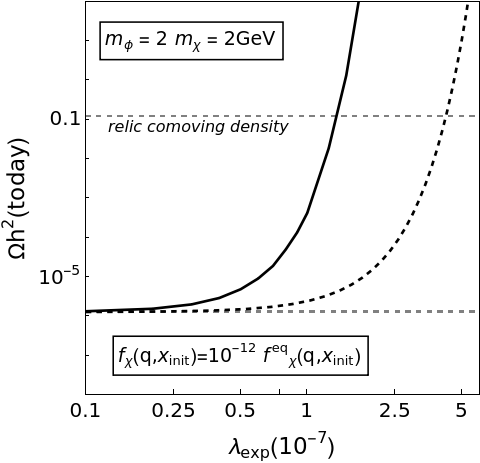}
    \caption{We plot the behaviour of dark matter comoving number density and dark matter abundance with $x$ and $\lambda_{\rm exp}$ in the left and right panel, respectively. 
    The solid lines are obtained by solving the unintegrated Boltzmann equation, while the dashed lines are obtained using the scaled equilibrium behaviour for distribution function as discussed 
    in appendix~\ref{sec:appendix}. For the purpose of the analysis, the masses of dark matter and bath particle are chosen to be 1 and 2 GeV, respectively.}
    \label{fig:eq-distribution}
\end{figure}

In figure~\ref{fig:eq-distribution}, we plot the differences in the abundances of dark matter obtained by directly solving the unintegrated Boltzmann equation and those obtained using the equilibrium assumption, as in reference~\cite{Bringmann:2021tjr} and illustrated in appendix~\ref{sec:appendix}. We observe that the couplings required to obtain the correct dark matter abundance differ in both cases. In particular, slightly lower couplings are needed to achieve the correct abundance in the numerical case. In the left panel of figure~\ref{fig:eq-distribution}, we show the growth of the dark matter number density, and it is evident that the growth is not uniformly exponential throughout as different momentum modes grow differently with increasing $x$.  Although the differences in the dark matter abundance are significant for the same value of the coupling as can be seen from the right panel of figure~\ref{fig:eq-distribution}, this can be attributed to the rapid exponential growth. However, the ballpark values for the couplings required to set the correct dark matter abundance using both assumptions are close. Therefore, if we are interested in obtaining approximate regions that satisfy the dark matter abundance, solving using the assumption of a scaled equilibrium distribution is practical.
\subsection{Case b: Out-of equilibrium decay of a non-relativistic particle}
\label{sec:inflation}
\begin{figure}
    \centering
    \includegraphics[width=0.475 \linewidth]{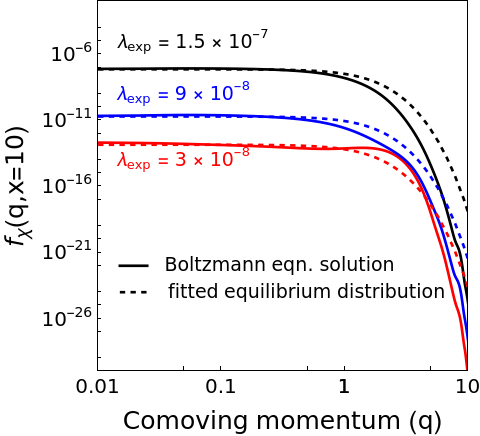} 
    \includegraphics[width=0.45 \linewidth]{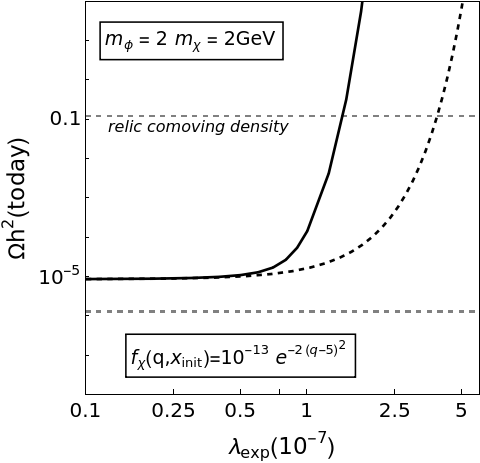}
    \caption{In the left panel, we plot the dark matter distribution after the exponential growth has ceased at $x=10$ for different couplings. Here, the solid lines correspond to the results obtained after solving the unintegrated Boltzmann equation, and dashed lines represent the fitted scaled equilibrium distribution. In the right panel, we show the dark matter abundance as a function of the coupling $\lambda_{\rm exp}$. The solid lines here as well correspond to the results obtained after solving the unintegrated Boltzmann equation, and the dashed lines are the results after solving the Boltzmann equation using scaled equilibrium approach, as discussed in the reference~\cite{Bringmann:2021tjr}. For the purpose of the analysis, the masses of dark matter and bath particle are chosen to be 1 and 2 GeV, respectively.} 
    \label{fig:decay-distribution}
\end{figure}

To illustrate the dependence on the initial conditions, 
as a second example, we consider the initial production 
of dark matter, at some epoch $x_{\rm init}$, from the non-thermal decay of a non-relativistic scalar particle ($\Phi$). 
We consider its mass to be much greater than the mass of the dark matter particle. The initial distribution function of dark matter can be determined by solving the Boltzmann 
equation which in this case is also a simplified equation because of negligible inverse process $\chi^* \chi \to \Phi$. 
The solution of the Boltzmann equation for dark matter depends on
the distribution function of $\Phi$ i.e. $f_\Phi$ or in other words on the details of $\Phi$ production
in the early universe~\footnote{The particle $\Phi$ can produced during the period inflation~\cite{Kofman:1994rk} or after inflation. 
It can be coupled with the thermal bath at the initial epochs or always remained 
thermally decoupled (for ex., see reference~\cite{Ballesteros:2020adh}.)}.
However, since we are considering the scenario where $m_\chi \ll m_\Phi$, the momentum of the dark matter 
produced through decays of $\Phi$ peaks roughly around $m_\Phi/2$.
Owing to this simplification, it is possible to simplify the analysis and
heuristically parametrise the distribution function of the dark matter independent of the details of the 
production of $\Phi$~\cite{Ghosh:2022hen} as
\begin{equation}
      f_\chi(q,x) = B  \; \exp{\bigg[-\frac{\left(q - q_{0} \right)^2}{2 \sigma^2}\bigg]} \; \Theta\left( x - x_{\rm init }\right) \;.
      \label{eqn:distributionfun-decay}
\end{equation}
Here $q_{\rm init} = {m_{\Phi}}/{\left( 2\; [s(x_{\rm init})]^{1/3}\right)}$ and $B$ and $\sigma$ are parameters which depends on the 
specific details of $\Phi$ production and its decays to the dark matter. We consider $\sigma$ to be small such that the distribution 
is peaked around $q_0$ in our computation. 
As before, the initial dark matter density $Y_\chi(x=x_{\rm init})$ obtained from 
decay of $\Phi$ is assumed to be much 
smaller than its abundance today $Y_\chi(x=x_{\rm today})$, 
with exponential growth mechanism as the dominant source of dark matter abundance.
Note that the decay of $\Phi$ is considered to be instantaneous in our analysis. 
With the initial distribution function in eqn.~(\ref{eqn:distributionfun-decay}), we 
can solve the unintegrated Boltzmann equation to compute the final dark matter distribution function. 
We show our results in figure~\ref{fig:decay-distribution} for $q_0 = 1.5$ and $\sigma=0.5$. 

It can be seen that from the left panel in figure~\ref{fig:decay-distribution}, the relic distribution function for the smaller couplings peaks 
at $q_{\rm init}$. For larger couplings, as exponential production kicks in, the distribution function develops non-zero values for low momentum. For larger couplings and lower comoving momenta, the distribution function behaves as if it is in equilibrium, with departures seen at higher comoving momenta and lower couplings. This result is similar to that obtained in section~\ref{sec:scaledeq},
but with different initial conditions. Although the initial dark matter distribution differs from 
the scaled equilibrium one, with the exponential growth mechanism, it starts to develop 
equilibrated behaviour for low comoving momenta and higher couplings. This can again be understood from the concept of the mean free path. 
In the right panel of figure~\ref{fig:decay-distribution}, we compare the results obtained using the numerical calculation versus 
approximate equilibrium approach. The obtained result is similar to the one obtained in section~\ref{sec:scaledeq}. 

Note that although our treatment of this section is simplified in terms of the choice 
of the initial distribution function, it does highlight the essence that the obtained dark matter distribution function . 

\textbf{Observation:} The behaviour of the dark matter distribution function after exponential growth appears to be general and independent of the initial distribution function, particularly for lower comoving momenta. While the exponential growth is inherently a non-thermal process, it demonstrates a degree of independence from the initial distribution function over a significant range of comoving momenta.

\section{Summary and concluding remarks}
\label{sec:summary}
In this manuscript, we determine the 
dark matter distribution function for the exponential growth mechanism 
by solving the unintegrated Boltzmann equation. We perform the analysis 
for the scenario where dark matter field is a complex scalar charged under 
$Z_3$ symmetry. The dark matter interactions with the scalar particle $\phi$ 
of the form $\chi^3 \phi$ triggers the exponential growth. The production 
continues until the process is Boltzmann suppressed. Note that 
this mechanism necessarily requires a finite non-zero number density of 
dark matter to be present in the early universe which can arise due 
to several initial processes -- like gravitational production, 
decay of the heavy particle, freeze-in etc.. 
While all these initial processes are capable of generating full 
dark matter abundance, in this manuscript however we assume that these process 
just seeds the small non-zero abundance and the dominant production of dark matter
occurs through the exponential growth process. 

The small ratio of dark matter number density to equilibrium number density simplifies the non-linear Boltzmann equation, transforming it into a linear coupled integro-differential equation.
We solve for the final distribution of dark matter using the unintegrated 
Boltzmann equation. This is in contrast with the previous works in the 
literature~\cite{Bringmann:2021tjr,Hryczuk:2021qtz,Bringmann:2022aim}, the results are obtained using the 
simplified assumption of dark matter phase space tracing the equilibrium 
distribution. In those cases, solving either the zeroth moment of the Boltzmann
equation, or coupled Boltzmann equation is sufficient. We solve the unintegrated form 
of the Boltzmann  equation with two choices of initial conditions. 
For case a, we make a simplistic choice 
that the initial dark matter distribution at $x=x_0$ is an accidental scaled equilibrium distribution. 
This choice helps to contrast with the results of the previous works in the
exponential growth scenarios~\cite{Bringmann:2021tjr,Hryczuk:2021qtz,Bringmann:2022aim}.
For case b, we consider that initial dark matter is produced by the out of equilibrium 
decay of a non-relativistic particle, and the distribution function 
of the dark matter is peaked around half the mass of the particle.

We find that the phase space distribution, obtained by directly solving the Boltzmann equation for two different cases, exhibits behaviour that is largely independent of initial conditions. For low values of the comoving momentum, the dark matter distribution begins to scale as the equilibrium distribution function, with this behaviour becoming more pronounced for larger coupling values. At higher values of the comoving momentum, the distribution deviates from equilibrium. It is important to note that the equilibrium behaviour here refers specifically to kinetic equilibrium. The observed behaviour of the distribution function suggests a degree of initial condition independence at lower comoving momenta. This result can be understood through the concept of the mean free path, which is shorter for larger couplings and low comoving momenta. Additionally, the transition to equilibrium behaviour is more pronounced with increasing couplings, due to the exponential dependence. A similar phenomenon is expected when transitioning from freeze-in to freeze-out, though with a slower increase, as shown in the appendix~\ref{sec:appendix}. Note that our analysis was conducted for specific choices of dark matter and bath particle masses. However, the conclusions drawn are general and applicable across a broad range of mass values, provided we remain in the limit where
$\phi$ does not directly decay to dark matter particles, and does not violate the bounds from big bang nucleosynthesis~\cite{BBN}. 

In summary, while the dark matter distribution does not reach full equilibrium across the entire momentum range due to the coupling required for the correct abundance being lower than that needed for kinetic equilibrium, equilibrium behaviour is still achieved in low-momentum modes. Consequently, although the assumption of dark matter following the equilibrium distribution during the exponential production phase is not perfectly precise, it offers a reasonable approximation for the purposes of this analysis. Given the computational complexity of solving the unintegrated Boltzmann equation, the use of the scaled equilibrium assumption for determining dark matter abundance remains a practical and effective approach to obtain ballpark results. For more accurate results, owing to the exponential sensitivity, one needs to solve the unintegrated Boltzmann equation.

There are several directions in which the present work can be extended.
We can consider the effects of other non-thermal production mechanisms 
which generate the initial dark matter abundance and find the correlation
with the already considered examples regarding the initial condition 
independence of the final dark matter distribution for low comoving momenta
and high couplings. 
In our work, we considered
$\phi$ to be coupled to a standard model bath. The results will 
change if $\phi$ was coupled to a hidden sector bath~\cite{Hryczuk:2021qtz,Bringmann:2022aim}, 
or was a non-thermal particle~\cite{Belanger:2020npe}. It would be interesting 
to work out these cases. 
It will be also interesting to work out the lower mass bounds on the dark matter from 
Lyman-alpha.

\vspace{0.2 in}
\section*{{Acknowledgements}} 
We would like to acknowledge the anonymous referee for his/her valuable suggestions. We would 
further like to thank Genevieve Belanger, Enrico Bertuzzo, Deeptak Biswas, Renata Zukanovich Funchal, Fernanda Lima Matos, Sreerup Raychaudhuri 
and Vinay Vaibhav for helpful discussions. We would also like to thank the fruitful discussions 
at the journal club seminars of IMSc and IFUSP, where part of this work was presented. The computational resources at
IFUSP were instrumental in performing this analysis. We would also like to 
acknowledge the participation of Shivam Gola in the preliminary stage 
of this work. The work is funded by the fellowship support from FAPESP under contract 2022/04399-4.

\appendix 

\section{Equilibrium assumption approach}
\label{sec:appendix}

In this appendix, we solve for the dark matter abundance assuming that the distribution function of the dark matter 
follows the scaled equilibrium behaviour. Substituting eqn.~(\ref{eqn:distribution-fun}) in the collision term of eqn.~(\ref{eqn:collision-operator}),
and computing the zeroth moment of the Boltzmann equation, we obtain:

\begin{eqnarray}
 \frac{1}{a^3} \frac{d \left( n_\chi a^3 \right)}{dt} &=&  \textcolor{red}{-} \frac{n_\chi(t) n_\phi^{\rm eq}(t)}{S n^{\rm eq}_\chi(t) n_\phi^{\rm eq}(t)} \int \frac{d^3 p_1}{(2\pi)^3 2 E_{1}}     \frac{d^3 p_2}{(2\pi)^3 2 E_{2}} \times \textcolor{blue}{f^{\rm eq}_\chi(p_1, t)} f^{\rm eq}_\phi(p_2, t)  
   \; \nonumber \\ 
  && \times \int \frac{d^3 p_3}{(2\pi)^3 2 E_3}  \frac{d^3 p_4}{(2\pi)^3 2 E_{4}}   (2\pi)^4 \delta^4(p_1 + p_2 -p_3 -p_4)  \times  |M|^2_{\chi \phi \to \chi^\dagger \chi^\dagger}   \nonumber \\
  && \textcolor{red}{+2}  \frac{n_\chi(t) n_\phi^{\rm eq}(t)}{S n^{\rm eq}_\chi(t) n_\phi^{\rm eq}(t)} \int \frac{d^3 p_1}{(2\pi)^3 2 E_{1}}  \frac{d^3 p_2^\prime}{(2\pi)^3 2 E_{2}^\prime} \frac{d^3 p_3^\prime}{(2\pi)^3 2 E_3^\prime}  \frac{d^3 p_4^\prime}{(2\pi)^3 2 E_{4}^\prime}  
   \; \nonumber \\ 
  && \times  (2\pi)^4 \delta^4(p_3^\prime + p_2^\prime -p_1 -p_4^\prime)   \times |M|^2_{\chi^\dagger \phi \to \chi \chi} \times \textcolor{blue}{f^{\rm eq}_{\chi}(p_3^\prime, t)} f^{\rm eq}_\phi(p_2^\prime, t)\;   \;.
\label{eqn:collision-operator-3}
\end{eqnarray}
Simplifying the above equation leads to the familiar dark matter number density equation as given in reference~\cite{Bringmann:2021tjr} with the backward scattering terms neglected 
in the small coupling limit. 
\begin{equation}
\frac{1}{a^3} \frac{d \left( n_\chi a^3 \right)}{dt} = n_\chi(t) n_\phi^{\rm eq}(t)  \langle \sigma v \rangle_{\rm exp} \;,
\label{eqn:equilibriumnumber}
\end{equation}
where 
\begin{eqnarray}
\langle \sigma v \rangle_{\rm exp} &=& \int \frac{d^3 p_1}{(2\pi)^3 2 E_{1}}     \frac{d^3 p_2}{(2\pi)^3 2 E_{2}} \times \textcolor{black}{f^{\rm eq}_\chi(p_1, t)} f^{\rm eq}_\phi(p_2, t)  
   \; \nonumber \\ 
  && \times \int \frac{d^3 p_3}{(2\pi)^3 2 E_3}  \frac{d^3 p_4}{(2\pi)^3 2 E_{4}}   (2\pi)^4 \delta^4(p_1 + p_2 -p_3 -p_4)  \times  |M|^2_{\chi \phi \to \chi \chi}  
\end{eqnarray}
Considering the toy model in our case with $\mathcal{L} = \frac{\lambda_{\rm exp}}{3!} \chi^3 \phi$ + h.c., the thermally averaged 
cross-section is given as: 
\begin{eqnarray}
    \langle \sigma v \rangle_{\rm exp} &=& \frac{2\pi^2 T (2\pi)^6}{n_\chi^{\rm eq}(T) n_\phi^{\rm eq}(T) S} 
    \int_{\left( m_\chi + m_\phi\right)^2}^{\infty} \frac{\sigma_{\rm exp}}{\sqrt{s}} \left( s - (m_\phi-m_\chi)^2\right)
    \left( s - (m_\chi+m_\phi)^2 \right) K_1{\left(\frac{\sqrt{s}}{T}\right)} \; \text{where}\;, \nonumber \\
\sigma_{\rm exp} &=& \frac{|\lambda|_{\rm exp}^2 }{16 \pi \sqrt{s}} \frac{\sqrt{s-4m_\chi^2}}{    \left(s-\left(m_\phi-m_\chi\right)^2\right)^{1/2}  \left(s-\left(m_\phi + m_\chi\right)^2\right)^{1/2}} \;.
\label{eqn:RG}
\end{eqnarray}
The number density Boltzmann equation in eqn.~(\ref{eqn:equilibriumnumber}) can be re-written in terms of comoving coordinates as:
\begin{eqnarray}
   \frac{d Y(x)}{d x} &=&  \frac{h_{\rm eff}(x)}{x H(x)}  Y_\chi n_\phi^{\rm eq} \langle \sigma v\rangle_{\rm exp} \;.
 \label{eqn:nodensity}
\end{eqnarray}
It can be seen that the comoving number density increases exponentially with $x$ until the process becomes Boltzmann suppressed.

\begin{figure}
    \centering
    \includegraphics[width=0.5\linewidth]{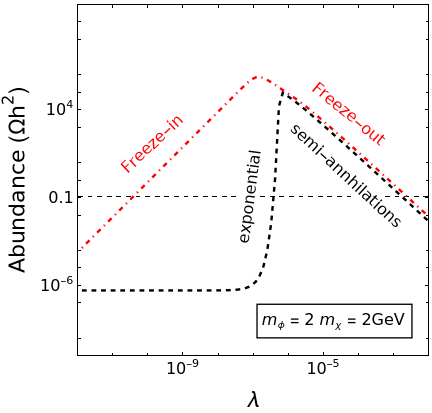}
    \caption{The phase diagram displaying complementarity between exponential production and semi-annihilations (in black dashed), along with freeze-in and freeze-out (in red dot-dashed). The horizontal black dashed line corresponds to the observed 
    dark matter abundance~\cite{Planck:2018vyg}}
    \label{fig:phase}
\end{figure}

Note that for exponential growth, the thermal averaged formula takes 
into account the mass differences of the colliding particles $\phi$ and $\chi$, 
that is why the obtained formula is slightly different from the conventional case~\cite{Gondolo:1990dk}.
We have used eqn.~(\ref{eqn:equilibriumnumber}) to obtained the dashed lines in the third panel of figure~\ref{fig:eq-distribution}. 

As mentioned earlier, the exponential production mechanism is a complimentary mechanism to 
dark matter production via semi-annhilations just like freeze-in and freeze-out are complimentary to one another. 
The former is active in the weak coupling region, while the latter becomes active in the strong coupling regime. 
We illustrate this complementarity in the phase diagram in figure~\ref{fig:phase} which illustrates rapid transition between exponential production and the semi-annihilation mechanism. 
In contrast, the transition between freeze-in and freeze-out is smooth\footnote{Note that for the freeze-in/freeze-out case ($\phi \phi \to \chi \chi^*$), the thermally averaged cross-section is given as:
\begin{eqnarray}
    \langle \sigma v \rangle_{\rm fio} &=& \frac{2\pi^2 T (2\pi)^6}{n_\chi^{2,\rm eq}(T) S} 
    \int_{4 m_\chi^2}^{\infty} \sigma_{\rm fio}{\sqrt{s}} \left( s - 4 m_\chi^2\right) K_1{\left(\frac{\sqrt{s}}{T}\right)} \;  
\bigg[\text{where} \; \sigma_{\rm fio} = \frac{|\lambda|_{\rm fi}^2 \sqrt{s-4m_\phi^2}}{ 16 \pi s \sqrt{s - 4 m_\chi^2}     } \bigg] \;.
\label{eqn:FI}
\end{eqnarray}}. It is because, dark matter abundance in freeze-in case is proportional to $\lambda^2$, in 
freeze-out it goes as $1/\lambda^2$. However, in exponential production scenario, dark matter abundance is 
proportional to $e^{\lambda}$ and grows very fast, while for semi-annihilation, the scaling dependence is similar to the freeze-out case. 
The differences can be seen in figure~\ref{fig:phase}. The correct relic abundance for exponential production is highly sensitive to the coupling. A slight increase in the coupling can cause dark matter to thermalize with the bath, shifting its production from the exponential mechanism to semi-annihilations. Due to this dependence, we observe that dark matter abundance for couplings where 
relic density is satisfied behaves almost like it's in equilibrium for low momentum values (see left panel of figure~\ref{fig:differentq} and right panel of \ref{fig:decay-distribution}). The mean free path in these cases is small, resulting in a near equilibrated behaviour. Note that for the phase diagrams plotted in figure~\ref{fig:phase}, we have considered both the forward and backward scattering terms, as in large 
coupling regions where equilibrium is attained both forward and backward scatterings are important. The transition region between freeze-in and freeze-out and correspondingly exponential production and the semi-annihilations lead to much larger abundances of the dark matter. To obtain the correct relic abundance, the coupling either has to be very small, or reasonably high. Note that the figure~\ref{fig:phase} is made to understand the qualitative differences and complementarity of the different kinds of dark matter production mechanisms. The phase diagram can be a bit different depending upon the presence of other mechanisms like self interactions of dark matter etc~\cite{Chu:2011be}.

\bibliographystyle{JHEP}
\bibliography{biblio}

\end{document}